\documentstyle[12pt]{article}
\begin{document}

\begin{center}

{\huge Statistical Measures of Complexity for

\vspace{0.2 cm}

Strongly Interacting Systems}

\vspace{0.5 in}

{\Large Ricard V. Sole$^{1,2}$ and Bartolo Luque$^{3}$}

\vspace{0.5 in}

(1) Complex Systems Research Group

Department of Physics- FEN

Universitat Politecnica de Catalunya

Campus Nord, Modul B5, 08034 Barcelona, Spain

\vspace{.4 cm}

(2) Santa Fe Institute

1399 Hyde Park Road, Santa Fe, NM 87501, USA

\vspace{0.4 cm}
(3) Centro de Astrobiolog\'{\i}a (CAB), Ciencias del Espacio, INTA 

Carretera de Ajalvir km. 4, 28850 Torrej\'on de Ardoz, Madrid (Spain)

\end{center}

\baselineskip=5 mm

\vspace{1.5 cm}

\begin{abstract}

In recent studies, new measures of complexity for nonlinear systems have
been proposed based on probabilistic grounds, as the LMC measure (Phys.
Lett. A {\bf 209} (1995) 321) or the SDL measure (Phys. Rev. E {\bf 59}
 (1999) 2). All these measures share 
an intuitive consideration: complexity seems to emerge in nature close to
instability points, as for example the phase transition points characteristic
of critical phenomena. Here we discuss these measures and their reliability 
for detecting complexity close to critical points in complex
systems composed of many interacting units. Both a two-dimensional spatially 
extended problem (the 2D Ising model) and a $\infty$-dimensional
(random graph) model (random Boolean networks) are analysed . 
It is shown that the LMC and the SDL measures can be easily generalized
to extended systems but fails to detect real complexity.
                                            
\end{abstract}
                                            
\begin{center}

\vspace{1 cm}

{\large \bf Submitted to Physical Review E}

\end{center}

\newpage

\section{Introduction}

  The quantitative characterization of complexity in nature has received
considerable attention over the last decade [1-4]. Although there is no 
universal agreement about the definition of what complexity is, increasing 
evidence suggests that it often emerges close to marginal instability
points of some kind. Critical phase transitions are a particularly well
studied example [5]. In all these systems, some generic properties 
like self-similar structures, $ 1/f $ noise and maximum information 
transfer seem to emerge in a rather spontaneous way.

  In order to quantify the degree of complexity in a system, many
measures have been proposed [1,6-9]. Comparative classifications
have been made based on general characteristics like the type of 
partitions in phase space, statistical structure, etc [10,11]. In spite of 
the large number of proposed measures of complexity, it seems to be a matter 
of fact that these measures have to share 
some stringent conditions.

\section{The complexity measure LMC}
  
  Recently, L\'opez-Ruiz, Mancini and Calbet (LMC) 
presented a new measure based on a simple set of assumptions [12]. 
In short, let us assume that an arbitrary dynamical system has N accesible 
states which belong to a set 
$ \Sigma _{\mu} = \{ x_{i}^{(\mu )}; i=1,...,N \} $ and have an associated
probability distribution
$$ \Pi _{\mu} = \{ p_i(\mu ) = P[x=x_i(\mu )]; i=1, ..., N \} \eqno(1) $$
Here $ \mu $ stands for a given parameter  which  
allows to describe the transition from the ordered
(low $ \mu $) to the disordered (high $ \mu $) regimes. As usual, 
$ \sum_{i=1}^N P_i(\mu ) = 1 $ and $ p_i(\mu ) > 0 \ \forall i=1, ..., N $.
The LMC measure is based on a combination of two different quantities: (a)
the Boltzmann entropy,
$$ {\rm H}_{\mu } ({\bf P}_{\mu})= - \sum_{j=1}^N p_j(\mu )\log p_j(\mu ) \eqno(2) $$
where ${\bf P}_{\mu} \equiv (p_1(\mu), ..., p_N(\mu))$ 
and (b) the so-called disequilibrium $ {\cal D}_{ \mu } $, defined as:
$$ {\cal D}_{\mu }({\bf P}_{\mu}) = \sum_{j=1}^N \bigg(p_j(\mu ) - {1 \over N}\bigg)^2 \eqno(3) $$
Using (2) and (3) the LMC measure of complexity is defined by the functional:
$$ {\cal C}_{ \mu } \bigl({\bf P}_{\mu} \bigr) = 
{\rm H}_{\mu }({\bf P}_{\mu}) {\cal D}_{\mu } ({\bf P}_{\mu})\eqno(4) $$
The basic idea involves the interplay of two different tendencies: the
increase of entropy as the system becomes more and more disordered and the 
decrease in $ {\cal D}_{\mu } $ as the system approaches 
equiprobability/disorder. The 
previous definition can be easily generalized to the continuum, i. e. : 
$$ {\cal C}_{ \mu } = {\rm H}_{\mu } {\cal D}_{\mu } = 
- \Biggl( \int p_{\mu }(x)\log p_{\mu }(x)\,dx \Biggr)    
\Biggl( \int p_{\mu }^2(x)\,dx \Biggr) \eqno(5) $$
This measure gives small values for highly homogeneous (ordered) or 
heterogeneous (disordered) states but it must reach a maximum at some
 intermediate value $ \mu ^* $. The application of this measure 
to some well known 
nonlinear dynamical systems described by iterative maps gave consistent 
results. For example, the study of the logistic map 
$ x_{n+1} = \mu x_n (1 - x_n) $ showed that $C_{\mu}$ undergoes a
rapid increase for $(\mu-\mu_c) \rightarrow 0$ close to the critical
point defining the onset of the period-three window.

\begin{figure}[htb]
  \vspace{10 cm} 
\includegraphics{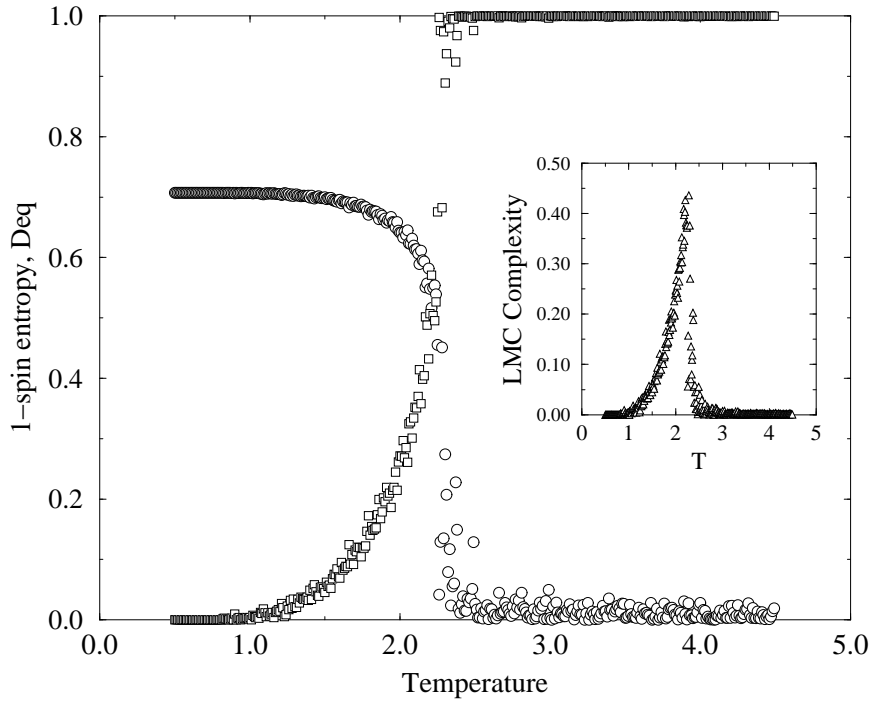}
\caption{Entropy (open circles) and disequilibrium (open squares) for a single spin 
    from a $40 \times 40$ Ising model. In all figures concerning this
    model, $\tau = 2\times10^4$ transients were discarded and averages
    were performed over $T=10^4$ time steps (Glauber dynamics). We can
    see that both quantities reach their extreme values near $T = T_c
    (=2.27)$. Inset: 1-spin LMC complexity (eq. 12) computed from the
    previous sets of values. A maximum is obtained at $T=T_c$ with a
    sharp decay to zero for $T>T_c$}
\end{figure}

  But there is no guarantee that the maximum of $ {\cal C}_{\mu } $ 
will occur at the appropiate $ \mu ^* $. In fact, the LMC measure
can be written as a combination of various $H_{\beta}$-Renyi entropies [4]
of a probability measure $\eta$. This seems reasonable, but there are too 
many possible ansatzes which would do the same. These problems have been 
recently discussed [13] in relation with 
the LMC measure. In this context, Feldman and Crutchfield [13] have shown that
the LMC measure is neither an intensive nor an extensive thermodynamic
variable. Another problem in the previous definition is the specific form
of $ {\cal D}_{\mu } $. The equiprobability is a consequence of an underlying 
assumption of weak or no interaction. Such distribution can be reached 
from a variational argument. Specifically, the variation of the functional:
$$ \delta \biggl[ {\rm H}_{\mu } - 
\alpha \bigl( \sum_{j=1}^N p_j - 1 \bigr) \biggr] = 0 \eqno(6) $$
leads to $ p_j = 1/N $ and so a maximum $ {\rm H}_{\mu} = \log N $. But if
other constraints play a role (as it occurs in 
far-from-equilibrium systems) then disorder does imply 
equiprobability.

\begin{figure}[htb]
  \vspace{10 cm}
  \includegraphics{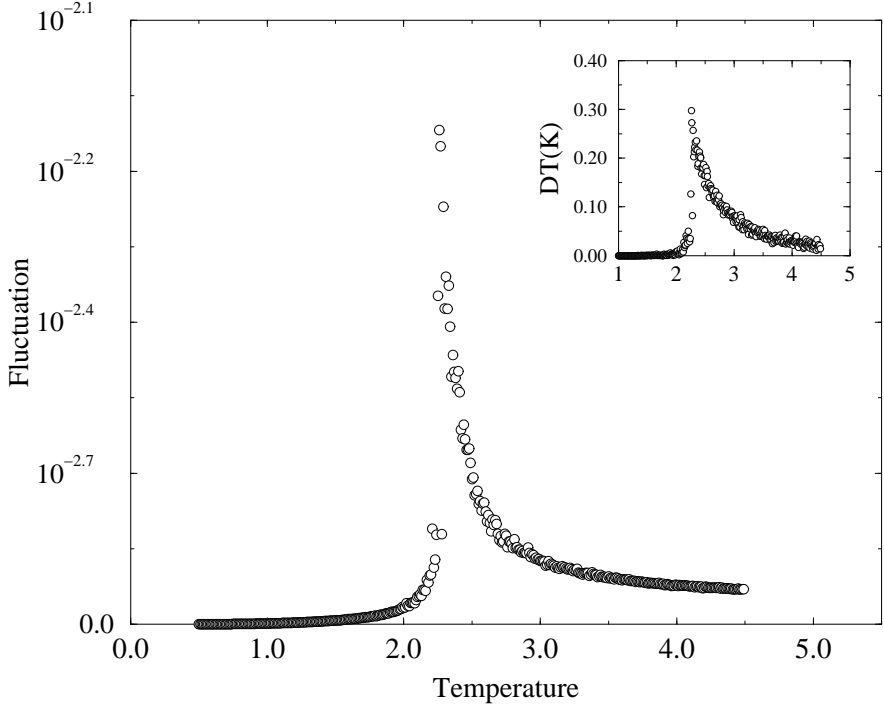}
  \caption{Average fluctuation of the magnetization $M$, defined from 
$\sigma(M) = \bigl < \vert 
M(t) - \bigl < M \bigr > \vert 
\bigr >$. A maximum is obtained at $T=T_c$, when fluctuations at all scales 
are observed. Three snapshots of a 2D Ising simulation are shown for three different temperatures. Inset: distance to independence for two nearest spins, as
calculated from (18). We can see that it behaves closely to $\sigma(M)$}
\end{figure}

 Let us first show that LMClike measures can be easily extended to complex,
spatially extended systems. The presence of correlations/interactions
in a given system
 can be measured through the spatial correlation function [14]. For a given
 sequence $ {\bf S} = s_1 , s_2, \dots ,s_N $ when 
$ s_i \in \Sigma = \{a_{\alpha}\}$ (with $i=1, \dots, k $) it is given by 
$$\Gamma (\delta) = \sum_{\alpha} \sum_{\beta} a_{\alpha} a_{\beta}
 P_{\alpha \beta}({\delta}) - \Biggr ( \sum_{\alpha} P_{\alpha} \Biggr)^2
 \eqno(7)$$
Two-point correlation functions are used in statistical physics as a
quantitative characterization of phase transitions. In the thermodynamic
limit $\Gamma(\delta)$ is expected to diverge at criticality
or reach a maximum if the system size is finite. For one of these systems, 
both temporal and spatial patterns contain information about the
existence of correlations arising from the local interactions.

 As a standard example, let us consider the
2D Ising model. In this paper all of our results for the Ising model
are obtained with a $L=40$ lattice and using Glauber dynamics.
This example allows us to consider a binary ($N = 2$) set of states and so
the relation between $ {\rm H}_{\mu }$ and $ {\rm C}_{\mu }$ is guaranteed
to be univocal [12]. Here $\mu$
is the temperature $T$ and the set of possible states for each
unit is simply $\Sigma_T=\{\uparrow, \downarrow\}$, where the arrows stands
for spin up and down, respectively. The mutual information $M(S)$ for two 
$\delta$-neighbors spins clearly fulfil the requeriment of being low at 
the ordered and chaotic regimes and maximum at the transition point.
We have:
$$ M(\delta) = \sum_{i \in \{\uparrow, \downarrow \}} 
 \sum_{j \in \{ \uparrow, \downarrow \}} 
P_{ij}(\delta) \log  \Biggr ( { P_{ij}(\delta)
 \over P_i(\delta) P_j(\delta)} \Biggr)^2 \eqno(8) $$ 
Now at $T<T_c$ most of the spins are in the same direction (say up) 
so $P_{\uparrow, \uparrow} \approx 1$ and $P_i=P_j=P(\uparrow) \approx 1$
and so $M(\delta) \approx 0$. Here $\delta_{ab}$ is the Dirac delta.
When $T>T_c$ we have $P(i,j) \approx P_iP_j$ and so
$M(\delta) \rightarrow 0$ as $T$ grows. Information transfer becomes 
maximum at $T_c$ [4] as a consequence of long-range correlations. 
In fact it can be shown that -for binary sequences-
a formal relation between the correlation function and the mutual
information can be derived [14]:
$$ M(\delta) \approx {1 \over 2} 
\Biggl ( { \Gamma (\delta) \over P_0(\delta) P_1(\delta) } \Biggr )^2 \eqno (9)$$
which leads to the conclusion that $M(\delta)$ decay to zero at a faster rate
than $\Gamma(\delta)$. So if $\Gamma(\delta) \approx \delta^{-\beta}$ then
 $M(\delta) \approx \delta^{- 2 \beta}$. Beyond these specific
relations, any properly defined complexity measure should give
equivalent results. 

  Clearly, if only a single spin is used, we can write the previous
definitions for the disequilibrium and entropy as 
$$ D_T(P(\uparrow)) = \Biggr ( P(\uparrow) - {1 \over 2}  \Biggr )^2 +
 \Biggr ( ( 1 - P(\uparrow)) - {1 \over 2}  \Biggr )^2 \eqno(10)$$
$$ H_T(P(\uparrow)) = - P(\uparrow) \log P({\uparrow}) -
 ( ( 1 - P(\uparrow)) \log  ( 1 - P(\uparrow)) \eqno(11)$$
and the LMC measure will be given by
$$ {\cal C}_T (P(\uparrow)) = D_T(P(\uparrow))  H_T(P(\uparrow)) \eqno(12)$$

Our results are shown in figure (1). We can see that $H_T$ grows fast
when approaching $T_c=2.27$ from $T<T_c$. This is an expected result
as far as $<M(T=T_c)>=0$ and so on the average $P(\uparrow)=1/2$. On the other
hand if we take $D_T$ we can see that it drops to zero at $T_c$, as expected.
The LMC measure is shown in the inset. We see a fast growth of $C_T$ with a
sharp decay at the critical point. So this characterization gives
nearly null complexity for $T>T_c$. This result, however, does not
correspond with the analysis of the model by means of standard techniques.
As an example, we show in figure 2 the average fluctuation 
of the magnetization for the
Ising model. It shows a maximum at $T_c$ and a decay for $T>T_c$. Such
decay is not as sharp as the one given from the LMC measure, and it 
measures in fact the existence of correlations for $T>T_c$ which are not
taken into account by the 1-spin LMC quantity.

\section{LMC extended and distance \\
to independence}

  The LMC measure can be extended by considering the statistics
of {\em interacting} units. The reason of this choice is clear:
long range correlations are created through interactions between nearest
spins, as defined by the Ising Hamiltonian $H_{Ising} = -\sum J S_i S_j$,
 being $J$
the coupling constant. A first possibility is to use the statistics of
$K$-spin blocks. The joint entropy for a $K-$block spin system is
defined as
$${\rm H}_T^{(K)} = -  \sum_{i_1 \in \{\uparrow, \downarrow \}}  
\sum_{i_2 \in \{\uparrow, \downarrow \}} . . . 
\sum_{i_K \in \{\uparrow, \downarrow \}}
 p (i_1, i_2, ..., i_K) \log p (i_1, i_2, ... , i_K) \eqno(13) $$
and the corresponding disequilibrium will be given by
$${\rm D}_T^{(K)} = -  \sum_{i_1 \in \{\uparrow, \downarrow \}}  
\sum_{i_2 \in \{\uparrow, \downarrow \}} . . . 
\sum_{i_K \in \{\uparrow, \downarrow \}}
\Biggl ( 
 p (i_1, i_2, ..., i_K) - {1 \over 2^ K} \Biggr )^2 \eqno(14) $$

For $K=2$, two given spins (here we take two nearest neighbors)
are considered and  a set of joint probability measures can be used 
$ p_{ij} = P [ s_a = i ; s_b = j ] $
, i.e., the joint probability of
having the $a$-spin in state $i$ and the $b$-spin 
in state $j$. Again, 
we have a normalization condition $ \sum p_{ij} = 1 $. Because $ p_{ij} $
introduces correlations in a very general way, we can redefine 
the disequilibrium in terms of these values as:
$$ {\cal D}_T^{ij} =
 \sum_{i \in \{\uparrow, \downarrow \}} 
 \sum_{j \in \{ \uparrow, \downarrow \}} 
 \Biggr( p_{ij} - 
{1 \over 4} \Biggr)^2 \eqno(15) $$
and the joint entropy as:
$$H_T=
-\sum_{i \in \{\uparrow, \downarrow \}} 
 \sum_{j \in \{ \uparrow, \downarrow \}} 
  p_{ij} \log p_{ij} \eqno(16)$$  
These are simple extensions of the previous LMC approach but
now the interaction (and so the intrinsic correlations) between
parts of the system are taken into account. In figure 3 the previous
quantities are shown. We can clearly observe that both quantities
behave smoothly close to $T_c$. This behavior leads
to an LMC-complexity that shows again a maximum at $T_c$ but
with a slower decay for $T>T_c$. This is consistent with
our explicit consideration of correlations into the joint probabilities.

\begin{figure}[htb]
  \vspace{10 cm}
  \includegraphics{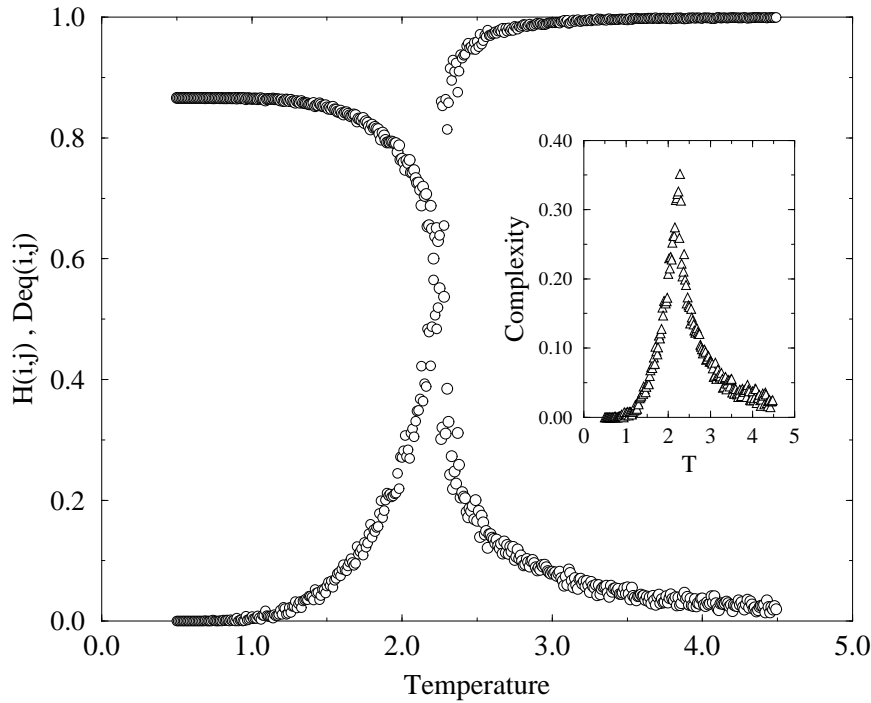}
  \caption{Joint entropy (16) (open squares) and disequilibrium (15) (filled triangles)
 for the 2D Ising model. Inset: corresponding LMC complexity (product of Joint 
entropy and disequilibrium). Now a maximum at $T_c$ is also obtained but 
correlations on both sides are detected (compare with figure 1, inset)}
\end{figure}

 An important drawback of these measures, as pointed before, is  the 
definition of the disequilibrium based on the distance to equal probabilities. 
We should remind that  for the $2D$-Ising model the entropy raises sharply
close to $T_c$ towards its maximum value, But this is not the case
in most problems. A simple generalization can be obtained by taking into 
account a much more natural measure, i. e. the distance to {\it independence}
defined, for a $K$-block as:
$$D_T^{(K)} = -  \sum_{i_1 \in \{\uparrow, \downarrow \}}  
\sum_{i_2 \in \{\uparrow, \downarrow \}} . . . 
\sum_{i_K \in \{\uparrow, \downarrow \}}
\Biggl (  p (i_1, i_2, ..., i_K) - \prod_{j=1}^K p(i_j) \Biggr)^2 \eqno(17) $$
which, for $K=2$ simply reads
$$ {\cal D}_T^{ij} = \sum_{i \in \{\uparrow, \downarrow \}} 
 \sum_{j \in \{ \uparrow, \downarrow \}} 
 \bigl( p_{ij} - 
p_i p_j \bigr)^2 \eqno(18) $$
Now, ${\cal D}_{\mu }^{ij}$ is a measure of how far are the subsystems 
from being independent (i.e., from the identity $ p_{ij} = p_i p_j $).
The distance to independence $D_T^I$, does not behave
 as $D_T^E$. It can be
easily shown that the distance to independence acts similarly to the
joint information (8) at both extremes (where total disorder leads to
$p_{ij}=p_ip_j$ or when a complete homogeneous distribution leads to
$p_{ii}=p_ip_i$ and zero for the other cases $j \ne i$). In fact:
$$M=\sum_{i,j}{p_{ij}\log{{p_{ij} \over p_ip_j}}}=-\sum_{i,j}{p_{ij}\log{\Biggl (1+{p_ip_j-p_{ij} \over p_ip_j}}\Biggr )} \eqno(19)$$
approximating to second order the logarithm, we have: 
$$M \approx -\sum_{i,j}{\Biggl( p_{ij} \Biggl ({p_ip_j-p_{ij} \over p_ip_j}
\Biggr )-{1 \over 2} \Biggl ({p_ip_j-p_{ij} \over p_{ij}}\Biggr )^2 \Biggr )}=
{1 \over 2} \sum_{i,j}{{\Bigl (p_{ij}-p_ip_j \Bigr )^2 \over p_{ij}}} \eqno(20)$$
i.e, the distance of the independence is a non-weighted approximation to the joint information.

 The behavior
of this measure for the 2D Ising model is shown in the inset of figure 2.
We see that $D_T^I$ behaves as the
fluctuations in the magnetization. This measure is able to detect the
onset of complex fluctuations (and the underlying correlations at
many scales) and decays for $T>T_c$ in the same way as the
average fluctuation. So in fact $D_T^I$ itself seems to be a simple
 and consistent measure of complexity very close to the underlying physics.
 As we will see below, this measure is also consistent in other cases.

\begin{figure}[htb]
  \vspace{8 cm}
  \includegraphics{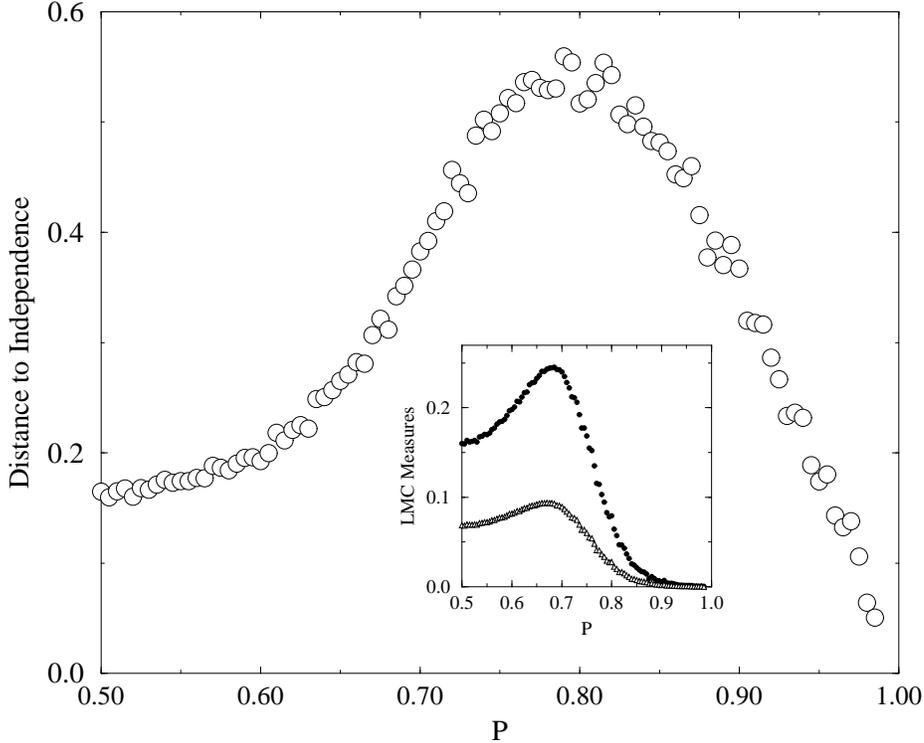}
  \caption{Distance to independence, $D_p^I$, as a
      function of the bias $p$, for RBN's of $K=3$ and $N=250$ spins.
      $p$ has been varied in intervals of size $\Delta p = 10^{-2}$.
      After a transient of length $\tau=250$ is discarded, we measure
      over $250$ time steps. Each point is the average over $400$
      samples. The theoretical analysis gives a critical point at $p_c
      = 0.79$ (derived from (20)) which is revealed by the distance to
      independence (21).  The snapshots show particular examples of
      the dynamics for $p=0.6$ (chaotic regime), $p=0.79$ (transition
      point) and $p=0.9$ (ordered regime).  Inset: the corresponding
      LMC complexities for the single-spin (open triangles) and
      two-spin (black circles) measures under identities compute
      conditions.  A maximum at $p \approx 0.69$ is obtained}
\end{figure}

\section{Testing LMC and SDL whit\\
 Random Boolean Networks}
The previous results might suggest that the generalized LMC complexity 
is an appropiate measure of correlations. This intuition, however,
 is not supported from the analysis
of other complex systems. As an example, let us consider the dynamics
of a random Boolean network (RBN) [15]. This model is defined
by a set of $N$ discrete maps: 
$$ S_i(t+1) = \Lambda_i (S_{i_1}, S_{i_2}, ... , S_{i_K} ) \eqno(21)$$
where $S_i \in \Sigma = \{0,1 \}$;  $i=1,2,..,N$ and 
$\Lambda_i (S_{i_1}, S_{i_2}, ... , S_{i_K} )$ is a Boolean
function randomly choosen from the set $\cal F_K$ of all the
Boolean functions with $K$ variables. So each spin in (21) receives
exactly $K$ inputs from a set of randomly choosen neighbors. This is
no longer a finite-dimensional extended system. Here the dynamics takes
place on a random directed graph. In fact, the statistical properties
of this model can be understood in terms of damage spreading on
a Bethe lattice [16]. 

It is well known that RBNs show a phase transition at
some critical points. Specifically, if 
$p=P[\Lambda_i (S_{i_1}, S_{i_2}, ... , S_{i_K} )=1]$ (i. e. $p$ is
the bias in the sampling of Boolean functions) a phase transition curve in the
$(K,p)$ plane is shown to exist [16-19] and is given by
$$ K = { 1 \over 2p(1-p) } \eqno(22)$$
The analysis of this system shows that at the critical 
point percolation of damage takes place for the first time i. e. a single flip
of a binary unit can generate an avalanche of changes through the whole system
and appropiate order parameters for the (second order) phase transition can
be defined.

 As argued in the previous example, maximum structure and correlation
should be expected at the critical point, and so maximum complexity.
In figure 4 we show the behavior of $D_p^I$ as a function of the
bias $p$ for a $K=3$ RBN. We can see a maximum at $p_c=0.79$, as predicted
by the critical condition given by (22). So the distance to the
independence  shows a maximum at the critical point, as expected from
a properly defined complexity measure. Here $D_p^I$ has been
computed by averaging
$$D_p^I = \Biggl < \sum_{j \in \{ i_1, i_2, ..., i_K\}} ( p_{ij} - p_i
p_j )^2 \Biggr >_N \eqno(23)$$
where $<...>_N$ stands for average over
all the units. We can similarly define the corresponding single-spin
and two-spin LMC measures for the RBN and the results are also shown
in the inset of figure 4. Althought both LMC measures have a maximum
at a given $0.5<p^*<1.0$, they fail to detect the critical
point. Here $p^*\approx 0.69 p_c$. This is understandable as far as
the $p$ parameter strongly influences (a priori) the statistical
distribution of state frequencies. 

In [20], Shiner, Davison and Landsberg introduce a new measure of complexity (hereafter SDL measure):
$$C_{\alpha\beta}=H^\alpha(1-H)^\beta \eqno(24)$$
where $H$ is the normalized Boltzmann entropy.
The SDL measure also tries to satisfy the criteria of maximum complexity between order and disorder as in LMC (in fact the LMC measure can be viewed as an approximation to the SDL as shown in [20]). $H$ is interpreted as a measure of disorder and $1-H$ as a measure of order. The parametrization with $\alpha$ and 
$\beta$ allow to fit ad hoc according to the specific cases considered.

The SDL measure has been criticized in [21]: the 
SDL measure give the same value for systems structurally different but with 
identical $H$. This problem is severe in RBN's. The entropy $H$ in 
RBN's depends, exclusively on the bias $p$, thus nets with differents connectivities (and then with different structural dynamic) but identical bias show
identical $H$ and then identical $C$ curves. Since the maxima of complexity
 depend too on $K$, the SDL measure will fail: we can always fit $\alpha$
 and $\beta$ in different ways for each
connectivity $K$ in such a way that we can recover the maximum. The value of
the normalized entropy $H$ that maximize the SDL measure is $\alpha/(\alpha+\beta)$ [20]. 
But it is easy to see from (22) and fixed $K$, that:
$$ H=-p_c\log p_c -(1-p_c)\log(1-p_c) =\alpha/(\alpha+\beta) \eqno (25)$$
allows an infinite number solutions of values for $\alpha$ and $\beta$,
which is far from satisfactory. Similarly, we can define $H$ in blocks, but the problem persists.
   
\section{DISCUSSION}

 In summary, we have analysed the validity of the LMC approach as an
effective measure of complexity for systems composed by many
units in interaction. Two different (although standard) problems have
been considered. For the 2D Ising model, where a phase transition is known
to occur at the Curie temperature, the LMC measures showed a maximum 
close to $T=T_c$. A measure based on the statistics from a single spin
failed, however, to detect correlations for $T>T_c$ although such correlations
exist. The reason was the lack of information about interactions between 
nearest spins and the fact that $P(\uparrow) \rightarrow 1/2$ as the critical 
temperature is approached from below. This problem was solved by 
considering an extension of the LMC measure to a joint probability
distribution. Such extension was able to show a maximum at $T_c$ and
well-defined corrlations. A simple extension of the
disequilibrium function was proposed -the distance to independence $D^I$- as
an alternative measure of complexity. This quantity (which is in fact
a second-order mutual Renyi information)  was shown to consistently
detect the critical point for the 2D Ising model and RBNs with a lower
computational effort than LMC complexity. A direct
extension of the LMC measure to the random Boolean network model
was shown to fail in detecting maximum complexity at the critical point.
The reason for this result is clear
is we consider that both the entropy and the disequilibrium change smoothly
with the $p$ parameter. Such parameter strongly influences the statistics
of the numbers of units in one of the two states, and in so doing
it definitely influences the values of both functions. Although 
correlations can be in principle detected, the relevance of $p$ in defining
a priori the values of entropies and disequilibrium is very important.
These results confirm the intuition that the LMC and SML measures (and similar
quantities) will fail to detect real complexity.

\vspace{1 cm}

{\bf ACKNOWLEDGMENTS}

\vspace{0.5 cm}

The authors thank the members of the CSRG for useful discussions. 
We also thank David Feldman, Jim Crutchfield and Mark Newman for
help at different stages of this work. This work 
has been partially supported by a grant DGYCIT PB94-1195 (RVS) and by Centro de Astrobiolog\'{\i}a (BLS). The authors 
thank the Santa Fe Institute where part of this work was done.

\newpage

\section{References}

\vspace{0.2 cm}

\noindent
[1] B. A. Huberman and T. Hogg, Physica D 22 (1986) 376.

\vspace{0.2 cm}

\noindent
[2] C. Langton, Physica D 42 (1990) 12.

\vspace{0.2 cm}

\noindent
[3] J. P. Crutchfield and K. Young, Phys. Rev. Lett. 63 (1989) 105.

\vspace{0.2 cm}

\noindent
[4] R. V. Sol\'e, S. C. Manrubia, B. Luque, J. Delgado and J. Bascompte, 
Complexity 1 (4) (1995/96) 13; C. Beck and F. Schlogl, Thermodynamics of chaotic systems
(Cambridge U. Press, Cambridge, 1993)

\vspace{0.2 cm}

\noindent
[5] P. Schuster in: Complexity: metaphors, models and reality. Santa Fe
    Institute Studies in the Sciences of Complexity, Vol. XIX, (1994)
    pp 382-413.

\vspace{0.2 cm}

\noindent
[6] H. C. Fogedby, J. Stat. Phys. 69 (1992) 411

\vspace{0.2 cm}

\noindent
[7] J. P. Crutchfield and D. P. Feldman, Phys. Rev. E 55 (1997) 1239

\vspace{0.2 cm}

\noindent
[8] P. Grassberger, Int. J. Theor. Phys. 25 (1986) 907

\vspace{0.2 cm}

\noindent
[9] M. Gell-Mann and S. Lloyd, Complexity 2(1) (1996) 44

\vspace{0.2 cm}

\noindent
[10] R. Wackerbauer, A. Witt, H. Atmanspacher, J. Kurths and H. Scheingraber,
     Chaos, Solitons and Fractals 4 (1994) 133

\vspace{0.2 cm}

\noindent
[11] H. Atmanspacher, C. Rath and G. Wiedenmann, Physica A 234 (1997) 819

\vspace{0.2 cm}

\noindent
[12] R. L\'opez-Ruiz, H. L. Mancini and X. Calbet, Phys. Let. A 209 (1995) 321

\vspace{0.2 cm}

\noindent
[13] C. Anteneodo and A. R. Plastino, Phys. Let. A 223 (1996) 348; D. P. Feldman and
J. P. Crutchfield, Phys. Lett. A, 238 (1998)

\vspace{0.2 cm}

\noindent
[14] W. Li, J. Stat. Phys. 60 (1990) 823

\vspace{0.2 cm}

\noindent
[15] S.A. Kauffman, The Origins of Order (Oxford University Press,Oxford,1993)

\vspace{0.2 cm}

\noindent
[16] B. Luque and R.V. Sol\'e, Phys. Rev. E 55 (1997) 257

\vspace{0.2 cm}

\noindent
[17] R.V. Sol\'e and B. Luque, Phys. Lett. A 196 (1995) 331

\vspace{0.2 cm}

\noindent
[18] B. Derrida and Y. Pomeau, Europhys. Lett 2 (1986) 739

\vspace{0.2 cm}

\noindent
[19] H. Flyvbjerg, J. Phys. A: Math. Gen. 21 (1988) L995

\vspace{0.2 cm}

\noindent
[20] J.S. Shiner, Matt Davison and P. T. Landsberg Phys. Rev E 59(2) (1999) 1459

\vspace{0.2 cm}

\noindent
[21] J. P. Crutchfield, D. P. Feldman and C. R. Shalizi chao-dyn/9907001

\end{document}